\documentclass[sigconf]{acmart}

\AtBeginDocument{%
  }

\usepackage{listings}
\usepackage{xcolor}
\usepackage{graphicx}
\usepackage[table,xcdraw]{xcolor}
\usepackage{amsmath}
\usepackage{multirow}
\usepackage{float}
\usepackage{tablefootnote}

\begin{document}

\title{VersaPants: A Loose-Fitting Textile Capacitive Sensing System for Lower-Body Motion Capture}

\author{Deniz Kasap}
\email{deniz.kasap@epfl.ch}
\orcid{0000-0002-6318-1705}
\affiliation{
\institution{École Polytechnique Fédérale de Lausanne (EPFL)}
\country{Switzerland}}

\author{Taraneh Aminosharieh Najafi}
\email{taraneh.aminoshariehnajafi@epfl.ch}
\orcid{0009-0009-9406-3072}
\affiliation{%
  \institution{École Polytechnique Fédérale de Lausanne (EPFL)}
  \country{Switzerland}}
  
\author{Jérôme Paul Rémy Thevenot}
\email{jerome.thevenot@epfl.ch}
\orcid{0000-0003-0209-025X}
\affiliation{%
  \institution{École Polytechnique Fédérale de Lausanne (EPFL)}
  \country{Switzerland}}

\author{Jonathan Dan}
\email{jonathan.dan@epfl.ch}
\orcid{0000-0002-2338-572X}
\affiliation{%
  \institution{École Polytechnique Fédérale de Lausanne (EPFL)}
  \country{Switzerland}}

\author{Stefano Albini}
\email{stefano.albini@epfl.ch}
\orcid{0009-0009-1592-0225}
\affiliation{%
  \institution{École Polytechnique Fédérale de Lausanne (EPFL)}
  \country{Switzerland}}

\author{David Atienza}
\email{david.atienza@epfl.ch}
\orcid{0000-0001-9536-4947}
\affiliation{%
  \institution{École Polytechnique Fédérale de Lausanne (EPFL)}
  \country{Switzerland}}

\renewcommand{\shortauthors}{Kasap et al.}

\begin{abstract}
We present VersaPants, the first loose-fitting, textile-based capacitive sensing system for lower-body motion capture, built on the open-hardware VersaSens platform. By integrating conductive textile patches and a compact acquisition unit into a pair of pants, the system reconstructs lower-body pose without compromising comfort. Unlike IMU-based systems that require user-specific fitting or camera-based methods that compromise privacy, our approach operates without fitting adjustments and preserves user privacy. VersaPants is a custom-designed smart garment featuring 6 capacitive channels per leg. We employ a lightweight Transformer-based deep learning model that maps capacitance signals to joint angles, enabling embedded implementation on edge platforms. To test our system, we collected approximately 3.7~hours of motion data from 11 participants performing 16 daily and exercise-based movements. The model achieves a mean per-joint position error (MPJPE) of 11.96~cm and a mean per-joint angle error (MPJAE) of 12.3° across the hip, knee, and ankle joints, indicating the model’s ability to generalize to unseen users and movements. A comparative analysis of existing textile-based deep learning architectures reveals that our model achieves competitive reconstruction performance with up to 22 times fewer parameters and 18 times fewer FLOPs, enabling real-time inference at 42 FPS on a commercial smartwatch without quantization. These results position VersaPants as a promising step toward scalable, comfortable, and embedded motion-capture solutions for fitness, healthcare, and wellbeing applications.

\end{abstract}



\maketitle

\section{Introduction}

The emergence of smart textiles is revolutionizing the way everyday garments interact with the human body. By embedding sensing and computation directly into fabric, clothing evolves from a passive layer into an active medium for perceiving motion, physiology, and context. Such smart garments promise a new class of unobtrusive and comfortable wearables that can be seamlessly integrated into daily life.

Applications for these technologies can span a wide range: wellbeing monitoring \cite{hughesriley2024}, elderly care \cite{Yang2019}, sports and fitness tracking \cite{yang2024}, entertainment, and mixed-reality \cite{chan2022}. All these applications benefit from continuous and personalized sensing of human movement and activity. Among these domains, motion capture (MoCap) stands as a fundamental enabler of any custom application by standardizing the digital representation of human movement decoupled from the application itself.

Traditional approaches for motion capture typically rely on complex and costly multi-camera setups \cite{Parks2019}. These advanced systems often depend on markers attached to anatomical landmarks, which are tracked in three-dimensional space using cameras positioned at different locations. Although they offer high spatial accuracy, they suffer from several practical limitations for being expensive, requiring dedicated calibration and constraining the wearer’s freedom of movement to a laboratory. Markerless multi-camera approaches using RGB cameras \cite{Colyer2018, dual_camera, JIANG2022S17}, markerless single-camera based solutions \cite{wham, humans4d}, and depth-based solutions \cite{Haimovich2021} have been proposed to alleviate some of these issues; however, they remain sensitive to occlusions, viewpoint changes, and privacy concerns. Alternative wearable sensor modalities, such as inertial measurement units (IMUs) \cite{xsens} or flexible sensors \cite{flex_fullbody}, offer portable motion tracking but introduce their own challenges: cumulative drift, placement sensitivity, and discomfort from rigid or tightly coupled sensors \cite{xiao2018}.

Integrating sensing directly into everyday clothing offers an appealing alternative. Textile-based sensing, in particular capacitive fabric sensors, can capture subtle deformations and contact variations across body segments without requiring tight body attachment or external infrastructure \cite{Geißler2024}. Such garments preserve comfort and social acceptability while enabling long-term continuous monitoring in natural settings \cite{comfort}. However, achieving reliable motion capture from loose-fitting garments remains challenging, as the signals depend not only on body motion but also on fabric behavior, fit, and environmental noise.

While prior research on motion capturing with smart garments has primarily focused on upper-body sensing \cite{mocapose, seampose, fip, lip}, the lower body remains comparatively underexplored. Yet, the lower body plays a pivotal role in understanding human activity: it governs locomotion, balance, and gait, which are essential indicators for mobility assessment, rehabilitation, sports performance, and elderly care. Continuous monitoring of lower-body motion using comfortable, everyday clothing could therefore unlock powerful real-world applications that remain inaccessible to conventional wearable or camera-based systems.

In this work, we present VersaPants, a textile-based motion capture system embedded in loose-fitting pants that combines textile-based capacitive sensing with embedded deep learning for on-device motion estimation. To the best of our knowledge, this is the first study that targets lower body motion capture using a loose-fitting garment approach. VersaPants introduces a fully integrated design, from analog front-end circuitry and data acquisition to model inference on a low-power microcontroller, enabling real-time operation without external computation or connectivity. This edge-centric approach preserves user privacy, reduces energy consumption, and supports continuous, autonomous operation suitable for everyday use.

Our contributions are as follows:
\begin{itemize}
    \item We design loose-fitting smart pants with a multi-channel capacitive sensing architecture for lower-body motion capture.

    \item We develop a transformer-based lightweight deep learning model capable of regressing 3D body motion directly from textile sensor data.
    
    \item We design a lightweight deep learning model that is $17.6 \times$ more efficient than prior work, enabling the first real-time (>30 FPS) motion capture inference on an embedded smartwatch platform.
    
    \item We tested the system across 11 users and 16 movements, highlighting its robustness, wearability, and generalization.
\end{itemize}

The remainder of this paper is organized as follows: Section \ref{sec:2} discusses related work on motion capture and textile sensing. Section \ref{sec:3} details the system design, including hardware, data acquisition, and pants design. Section \ref{sec:4} describes the deep learning pipeline. Section \ref{sec:5} presents the experimental setup and evaluation. Section \ref{sec:6} discusses limitations and future work, and Section \ref{sec:7} concludes the paper.

\section{Related Work} \label{sec:2}

\subsection{Textile Capacitive Sensing for Motion and Activity Recognition}

Over the past decade, textile capacitive sensing has been widely explored for body-motion applications, enabling fabrics to detect deformation, stretch, or contact changes. Early studies primarily focused on classification-based tasks, such as gesture recognition \cite{mocapaci}, fitness or exercise classification \cite{gym_passive, seamfit}, and domain-specific human–computer interaction systems. A notable subfield involves smart gloves, where capacitive or hybrid sensing has been used for hand gesture classification \cite{glove_gestures}, drone control \cite{captainglove}, and even finger pose reconstruction \cite{Geißler2024}.

While these systems demonstrate the versatility of textile sensing, they address classification problems that involve tracking only a few actions. In contrast, motion capture involves continuous regression of body pose in three-dimensional space, a task that is inherently more complex. Unlike discrete classification, MoCap demands temporally smooth, spatially consistent estimations that reflect real-time human motion, requiring models to interpret high-dimensional, nonlinear, and dynamic sensor data.

\subsection{Wearable Motion Capture: From Tight-Fitting to Loose-Fitting Garments}

Wearable motion capture systems have traditionally relied on tightly coupled sensors to ensure accurate tracking. IMUs often dominate this category, being employed in several commercial and research systems such as Xsens, SIP, DIP, PIP, TransPose, and IMUPoser \cite{xsens, sparse_inertial, deep_inertial_poser, Transpose, PIPCVPR2022, IMUPoser}. Despite their success, these approaches require tight-fitting suits or rigid mounting to minimize drift and alignment errors, which compromise comfort and long-term usability \cite{xiao2018}.

To relax this constraint, several works have investigated loose-fitting sensing with IMUs. The LIP system \cite{lip} integrates four IMUs into a loose jacket and compensates for fabric motion through synthetic data simulation and autoencoder-based correction. Extending this idea, FIP \cite{fip} incorporates additional flex sensors at the elbows and leverages diffusion models to generate synthetic samples for sensor displacement calibration. These studies mark an important step toward everyday wearable motion capture; yet, their distributed circuitry and calibration overhead introduce significant complexity and reduce garment practicality, limiting factors such as washability and comfort in real-world use.

\paragraph{\textbf{Loose-Fitting Textile-Based Approaches}}

Recent work has begun to explore textile-based sensing for loose-fitting applications, aiming to retain the flexible properties of garments while capturing rich motion signals. The pioneering system MocaPose \cite{mocapose} demonstrated the feasibility of capacitive motion capture using a jacket with 16 conductive textile patches. A Convolutional Neural Network (CNN) maps the temporal variations in capacitance to upper-body joint positions, proving that capacitive sensing can reconstruct continuous motion.

Building on this foundation, SeamPose \cite{seampose} proposed a more integrated approach by embedding conductive threads within the seams of a long-sleeve shirt. Instead of large surface patches, these fine-grained textile conductors capture localized capacitance changes, which are then processed by a Long Short-Term Memory (LSTM) model to predict 3D joint rotations for a skinned human mesh. While these works represent essential milestones, they remain restricted to upper-body configurations, and lower-body continuous pose estimation with textile sensors is still largely unexplored.

\paragraph{\textbf{Lower-Body Motion Sensing}}

A few studies have attempted to monitor lower-body kinematics using textile-integrated sensors; yet, they predominantly rely on tight-fitting designs. For instance, \cite{tavassolian2020, Gholami2019} employ fabric-based strain or inductive sensors wrapped tightly around joints to monitor running dynamics and hip joint angles. Similarly, \cite{galli2023} presents a knee band with capacitive electrodes to measure continuous knee flexion. Another study, SoftCap \cite{soft_capacitive}, introduces soft capacitive elements attached directly to the skin for lower-body angle estimation, achieving a linear correlation between sensor outputs and joint angles.

While effective for laboratory evaluation, all these approaches depend on skin-tight coupling and localized sensor placement, diverging from the vision of everyday loose-fitting garments suitable for natural movement and continuous wear. Current lower-body systems either compromise comfort with tight-fitting designs or introduce excessive complexity through the use of deep neural networks. To bridge this gap, we propose VersaPants: a loose-fitting, textile-capacitive sensing system for lower-body motion capture.

\section{VersaPants} \label{sec:3}
This section presents the mechanical and electronic design of the proposed wearable system, VersaPants. The system comprises two main components: (1) a data acquisition unit (DAU) that continuously measures and transmits multi-channel capacitive signals in real time, and (2) the garment itself, where conductive textile patches are strategically positioned over lower-body joints.  

\subsection{Electronic Design}

The electronic design integrates the VersaSens main board from the open-hardware VersaSens platform~\cite{versasens} as the primary controller, along with a dedicated custom board designed as a capacitive sensing analog front end that converts capacitance measurements into digital data.

\paragraph{\textbf{VersaSens Main Board.}}

VersaSens is an extendable, modular, multimodal platform. Its flexible design supports multiple sensor modalities, hardware accelerators, and adaptable form factors, making it suitable for multipurpose edge-AI applications. 

The Main module features a dual-core Arm® Cortex-M33 nRF5340 SoC (Nordic Semiconductor®) comprising a processing core and a network core. It also integrates a power management unit responsible for regulating power supply and charging the battery, an SD card for data storage, and an IMU for inertial measurements.

\begin{figure}[h]
  \centering
  \includegraphics[width=\linewidth]{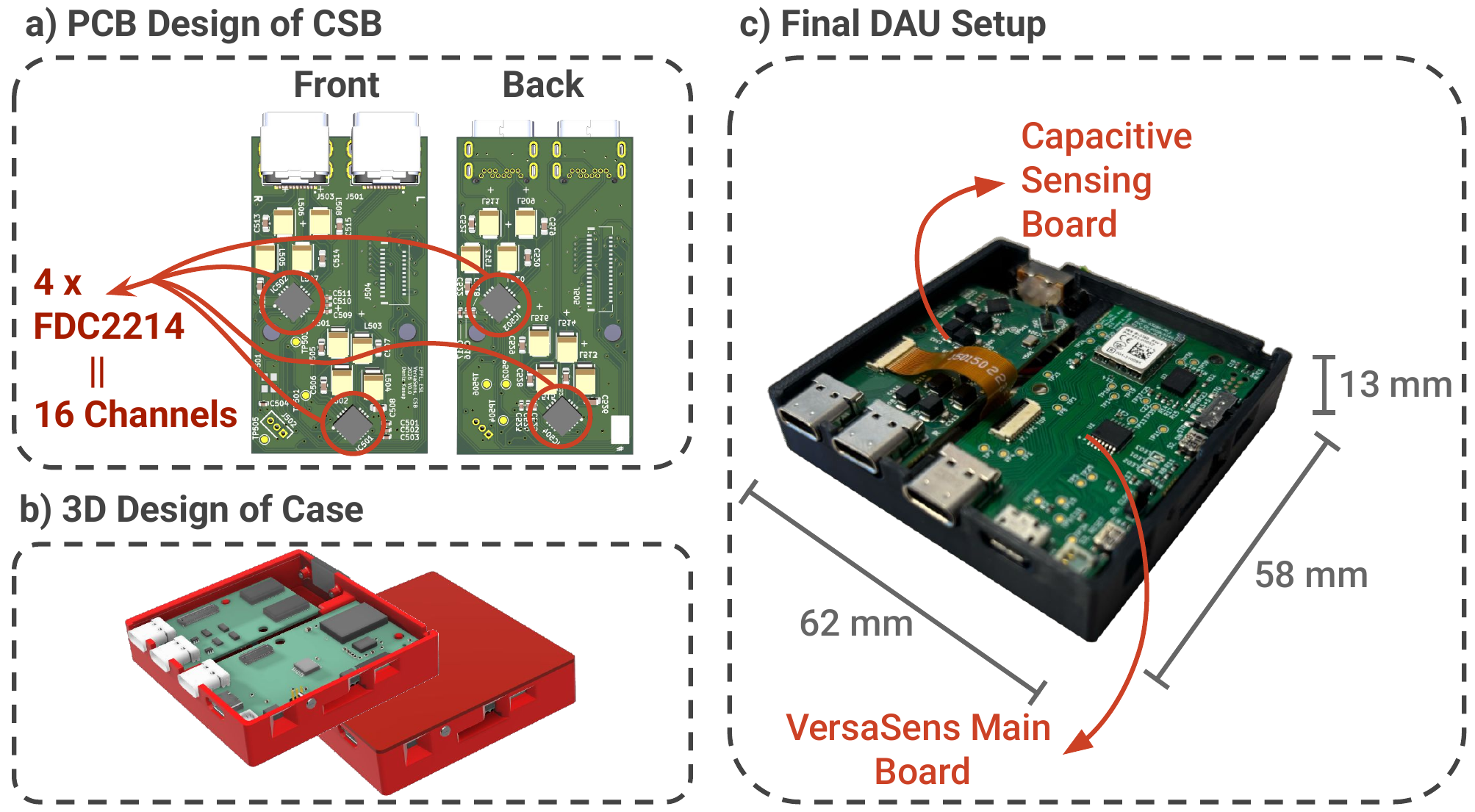}
  \caption{Design of Data Acquisition Unit: a) Capacitive Sensing Boards with 4 FDC2214 units, supporting up to 16 channels. b) 3D CAD design of the plastic enclosure. c) The final DAU with manufactured PCBs and printed casing.}
  \label{fig:dau}
\end{figure}

\paragraph{\textbf{Capacitive Sensing Board}}
Leveraging the multimodal capabilities of the VersaSens platform, we developed a custom Capacitive Sensing Board (CSB) to acquire capacitance measurements from each textile patch, ensuring seamless system integration.

The CSB maintains the same form factor and bus topology as VersaSens and incorporates four Texas Instruments FDC2214 frequency-to-digital converters\footnote{https://www.ti.com/product/FDC2214}, enabling the measurement of up to 16 capacitive channels.

FDC2214 is designed to be robust to cross-talk artifacts within its four channels, given the excitation frequencies of each channel (capacitance value) is distinct \cite{fdc}. To mitigate inter-chip crosstalk between four FDC2214 chips, a chip multiplexing strategy is employed, such that only one chip is active at a time. The nRF5340 core on VersaSens polls all FDC2214 devices at 30~Hz, aligning with typical monocular camera frame rates for multimodal synchronization. Acquired data are transmitted wirelessly via Bluetooth Low Energy (BLE) for real-time monitoring and logging.

 \begin{figure*}[h]
  \centering
  \includegraphics[width=\textwidth]{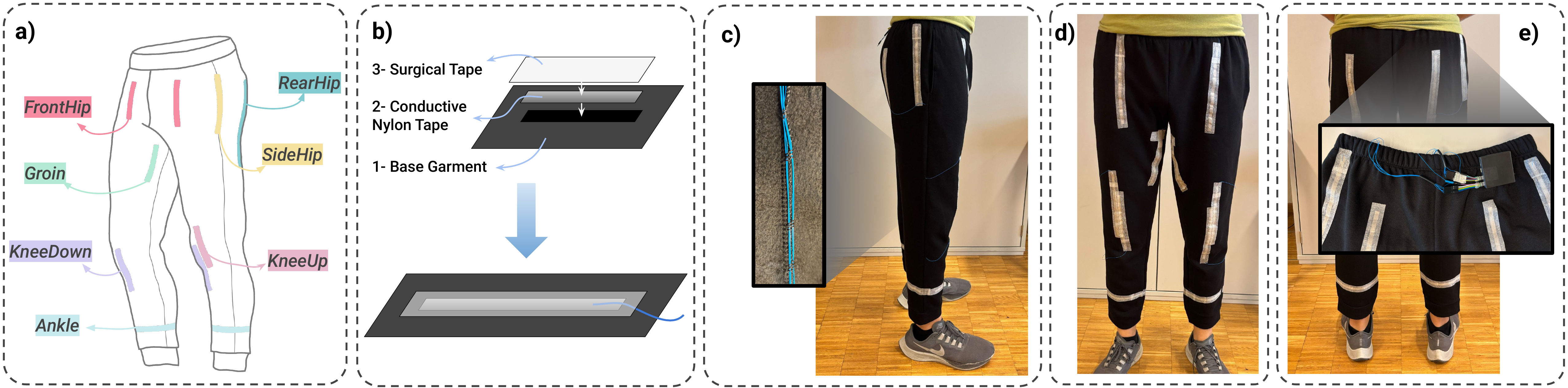}
  \caption{Production of VersaPants: a)Sensor placements for the pants. b) Layered construction of each channel. c) Side view of the pants with wires channeled through the inner seams. d) Frontal view of the pants. e) Rear view of the pants, with DAU attached to the back part of the waist.}
  \label{fig:pant_img}
\end{figure*}

\subsection{Wearable Design and System Integration} \label{sec:wearable_design}
This subsection focuses on garment design, considering the sensor placement and integration. The configuration of conductive patches follows the biomechanical structure of the lower body, accounting for the degrees of freedom (DoF) of each major joint.  

The hip joint, a ball-and-socket articulation, has three DoFs, rotation along the transversal, sagittal, and longitudinal axes, while the knee joint, a hinge joint, exhibits a single DoF along the transversal axis \cite{Arnold2009}. Our design strategy aimed to capture the complete kinematic nature of the lower body by placing at least one patch corresponding to each degree of freedom. This resulted in an intentionally over-constrained design of 7 patches per leg.

\paragraph{\textbf{Patch Placement Strategy.}}
In total, 7 patches per leg are integrated with the pants, with symmetric placement on both legs, yielding 14 channels in total, as illustrated in Figure \ref{fig:pant_img}.
\begin{description}
    \item[1-2] \textbf{\textit{FrontHip} / \textit{RearHip}:} Capture hip flexion and extension along the transversal axis.  
    \item[3] \textbf{\textit{SideHip}:} Capture hip abduction and adduction along the sagittal axis.  
    \item[4] \textbf{\textit{Groin}:} Measure inter-leg proximity, indicative of stance width. When the legs are close (e.g., standing at attention), the two groin patches exhibit high mutual capacitance.  
    \item[5-6] \textbf{\textit{KneeUp} / \textit{KneeDown}:} Capture knee flexion and extension. Two vertically offset patches per knee compensate for anatomical variation and garment looseness across users. Depending on leg length, either the upper or lower patch dominates signal response.  
    \item[7] \textbf{\textit{Ankle}:} Measure proximity between ankles, as well as between ankle and posterior thigh regions during knee bending.  
\end{description}

\paragraph{\textbf{Conductive Textile Patches.}}
For the conductive textile patches serving as single-ended sensing electrodes in the CSB, "Conductive Nylon Fabric Tape" from Adafruit was selected\footnote{https://www.adafruit.com/product/3960}. The tape consists of silver-coated nylon fibers providing sufficient conductivity, where we measured resistance below $2.5~\Omega$/m. Among several off-the-shelf conductive textiles tested, this material proved to be the most fabrication-friendly, owing to its integrated conductive adhesive layer that enables direct attachment to the garment without the need for additional bonding materials.

Given the loose-fitting nature of the pants, no elastic or stretchable properties were required for the conductive elements. The patches, with a thickness of approximately $8$~mm, minimize surface wrinkling over large areas such as the knee, and instead capture localized deformation along the dominant axis of joint motion. Each patch was cut to a length between $10$~cm and $30$~cm, depending on its placement, and securely adhered to the garment. Variable patch lengths were intentionally selected to produce distinct excitation frequencies across the channels, hence reducing potential crosstalk within and among FDC2214 chips.

\paragraph{\textbf{Garment Construction.}}
A commercial off-the-shelf pair of men’s jogging pants (58\% polyester, 42\% cotton, size L, W34L34, relaxed fit) was selected to ensure comfort and adaptability across users, complying with loose-fitting approach\footnote{https://www.decathlon.ch/en/p/pantalon-jogging-chaud-fitness-homme-100-noir/\_/R-p-307272?mc=8543999\&c=bleu}. Conductive patches are connected to the DAU via 32~AWG unshielded copper wires routed through the inner and outer seams of the pants to preserve flexibility and prevent damage during motion. The connection to DAU is made through two female USB Type-C connectors, grouping the channels on the left and right legs. Each patch was then bonded to its corresponding wire termination and reinforced using medical-grade surgical tape, chosen for its strong adhesion to textile substrates and biocompatibility. To ensure easy detachment and maintenance, intermediate male–female DuPont connectors were added inline before the USB connector interface.

\paragraph{\textbf{Data Acquisition Unit (DAU)}}
The DAU integrates the VersaSens main board and CSB within a 3D-printed enclosure fabricated from biocompatible polylactic acid (PLA) (Figure \ref{fig:dau}). The final casing measures 62~$\times$~58~$\times$~13~mm, weighs 38~g, and is powered by a 3.7~V, 500~mAh Li-Po battery, providing continuous operation for approximately 36 hours. The compact, lightweight form factor allows the DAU to be clipped unobtrusively to the back of the pants without compromising user comfort or movement.

\begin{figure*}[h]
  \centering
  \includegraphics[width=\textwidth]{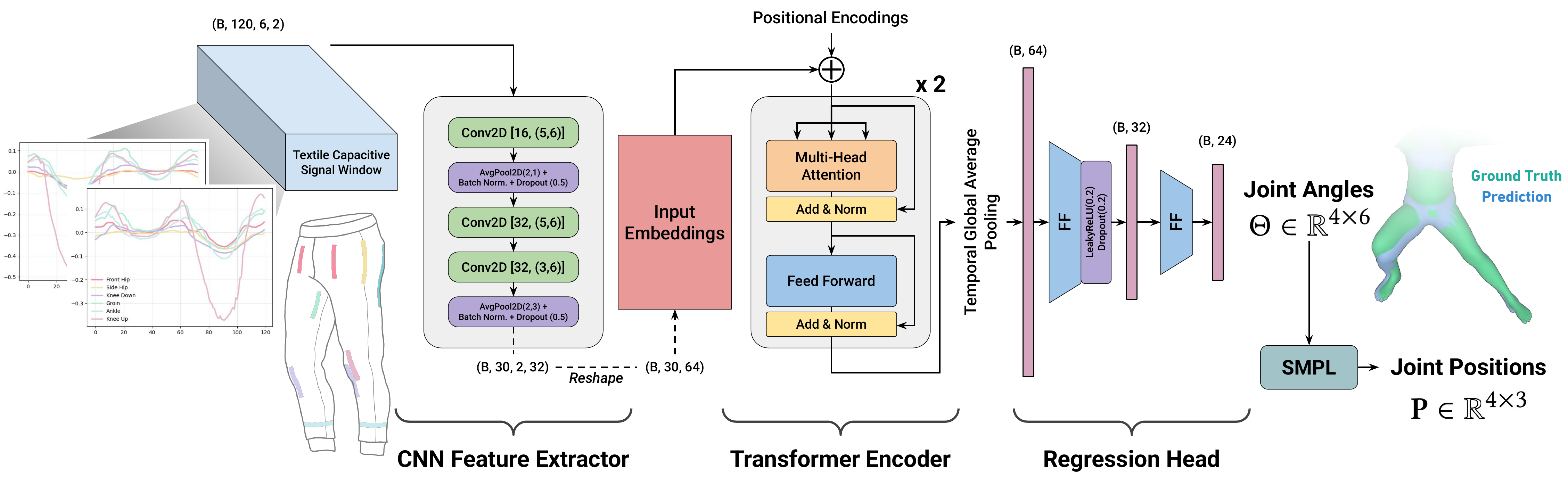}
  \caption{Deep Learning Pipeline for the Proposed CNN2D+Transformer Architecture.}
  \label{fig:model_arch}
\end{figure*}

\section{Motion Capturing from Textile Signals Concept}\label{sec:4}
Any deformation or displacement of the conductive textile patches alters the capacitance between the textile and the underlying skin, which can then be sensed in real-time by the custom capacitive acquisition board. The goal of this work is to learn a mapping between the temporal evolution of these capacitance signals and the corresponding lower-body joint angles. Given a time window of length $N=120$ samples (4 seconds at 30 Hz), the input to the network is a matrix $\mathbf{A} \in \mathbb{R}^{N \times K \times 2}$ that contain the time-series sensor values, where $K=7$ denotes the number of textile channels per leg, and the output is a vector of joint rotations $\Phi \in \mathbb{R}^{J \times 3}$ in three-dimensional space, with $J=4$ the number of relevant joints.

\subsection{Ground Truth Definition}

Accurate ground-truth joint positions and rotations are essential for supervised learning. However, traditional motion capture systems are impractical for loose-fitting garments, as markers move with the garment \cite{ray2023}. To overcome this limitation, a single-monocular video-based approach was adopted using the WHAM (World-grounded Humans with Accurate Motion) framework \cite{wham}. WHAM reconstructs accurate 3D human pose and global motion trajectories directly from video sequences. Image-extracted 2D keypoints with a Vision Transformer are lifted into 3D poses. Its output consists of temporally consistent "Skinned Multi-Person Linear" (SMPL) parameters, which provide a reliable ground truth in an efficient and markerless way.

\paragraph{\textbf{SMPL Model}}

The SMPL model \cite{smpl} is a parametric, skinned 3D human body representation widely adopted for pose regression and motion reconstruction. The model inputs a body shape $\beta \in \mathbb{R}^{10}$ and body pose $\boldsymbol{\theta} \in \mathbb{R}^{24 \times 3}$ parameters, yielding a triangular body mesh with $6890$ vertices. This model is used to visualize the body pose and to obtain the joint locations in 3D space (joint rotations). The SMPL framework captures both skeletal kinematics and soft-tissue deformations through learned blend shapes, resulting in anatomically realistic motion reconstructions \cite{smpl}.

For this study, only the lower-body joints are considered:
\begin{itemize}
    \item \texttt{left\_hip}, \texttt{right\_hip}, \texttt{left\_knee}, and \texttt{right\_knee} for rotational regression,
    \item \texttt{left\_knee}, \texttt{right\_knee}, \texttt{left\_ankle}, and \texttt{right\_ankle} for positional regression.
\end{itemize}

The rotations of the ankles are excluded as they are (1) not captured by the vision-based human pose reconstruction models, and (2) the movement of feet is not in the vicinity of the pants to produce significant textile signal variation. Estimation of global orientation (rotation of the hip joint) and body shape parameters of SMPL are not in the scope of this study.

\subsection{Deep Learning}

With 7 capacitive channels per leg and extracted SMPL ground truth poses, we design a neural model that learns the mapping from spatio‐temporal sensor patterns to human pose. Our design was driven by two objectives: (1) to achieve robust (for unseen users and movements) and accurate regression of body pose, and (2) to ensure that the model remains suitable for deployment on portable platforms (smartwatches and mobile devices) with tight limitations on memory, compute, and power consumption. Given an eventual target of on‐body real-time inference, the architecture must achieve high-quality reconstruction with minimal footprint.

State-of-the-art methods on motion capturing with loose-fitting textile capacitive sensing, such as MocaPose and SeamPose, employ different models. The first one uses an intensive CNN2D + CNN1D stack \cite{mocapose}, while the latter proposes a bidirectional LSTM \cite{seampose}.
While these architectures achieve reasonable joint estimation accuracy, they come with substantial computational and memory overhead due to the sequential nature of recurrent networks or the large convolutions required in deep CNNs.

In contrast, we adopt a Transformer-based architecture. Transformers \cite{attention} have achieved remarkable success in natural language processing \cite{tucudean2024natural}, but they have proven efficient also in the biomedical domain \cite{ma2023tsd, amirshahi2024metawears}, and even in sensor-based human pose estimation. For instance, \cite{TransformerInertialPoser} and \cite{IMUOptimize} employ Transformers for IMU-based pose estimation, while \cite{FabricBasedPressureSensor} combines pressure sensor arrays with IMUs for lower-limb motion capture.

Inspired by \cite{IMUOptimize, FabricBasedPressureSensor}, we base our architecture only on the encoder part of Transformers. Input embeddings are first extracted through a compact 2D convolutional network, which efficiently captures multi-channel spatial correlations among sensor inputs. Then, the Transformer encoder models long-range temporal dependencies across the sequence without the sequential overhead of LSTMs, enabling joint spatial-temporal learning while keeping the parameter count and FLOPs lower than purely recurrent or deep CNN architectures.

\paragraph{\textbf{Input \& Output Preprocessing}}
The FDC2214 chips in CSB output raw frequency numbers, before being converted to capacitance values \cite{fdc}. However, this latter step is not necessary, as the channel frequency is sufficient to describe the characteristics of motion. The raw frequency values of $\mathbf{A}$ are quite large, so they are normalized by scaling between the maximum $A_{max}$ and the minimum observed value $A_{min}$ in the recordings.

$$\mathbf{A}^{N \times K \times 2}_{ij} = {{\mathbf{A}^{N \times K \times 2}_{ij}} - A_{min}\over {A_{max} - A_{min}}} \quad \text{where} \quad 0 \leq i < N \ \ \text{and} \ \ 0 \leq j < K$$

In order to mitigate the variance among the subjects, we subtract the baseline values of each channel, collected at the beginning of the recording while the subject is stationary and standing still, from the whole recording. 

For output representation, SMPL's joint orientations are represented in axis-angle representation, $\mathbf{\phi^{4 \times 3}}$ for the 4 lower body joints. After the forward pass of SMPL, the joint positions $\mathbf{p^{4 \times 3}}$ are obtained. Also, each joint's axis angle representation can be converted into a rotation matrix $\mathbf{R} \in \mathbb{R}^{3 \times 3}$. For the deep learning applications, it has been shown that using the $6D$ representation of this rotation matrix better addresses the discontinuities in angle values, as they are bound between $[-\pi, +\pi]$ or $[0, 2\pi]$ \cite{6D_angle}. Therefore, the model outputs 6D representation of each joint, yielding a matrix $\mathbf{\Theta} \in \mathbb{R}^{4 \times 6}$ for four lower-body joints.

\paragraph{\textbf{Model Architecture}}
The complete architecture is illustrated in Figure~\ref{fig:model_arch}. Given a batch $B$ of windows of length $N$, where $K$ per-leg channels are collected separately, the input tensor with shape $(B, N, K, 2)$, is first processed by a sequence of three 2D convolutional layers with filter dimensions $(16, 32, 32)$ and linear activations. Average pooling, batch normalization, and dropout layers are applied after the first and third convolutional layers, producing a tensor of shape $(B, N/4, 2, 32)$. This tensor is reshaped to $(B, N/4, 64)$ and a learnable positional encoding is then added. Two layers of Transformer encoder, each with four attention heads, capture spatio-temporal relations through self-attention. Temporal global average pooling reduces the output to a tensor of shape $(B, 64)$, which is passed to a fully connected network to regress lower-body joint angles $\Theta \in \mathbb{R}^{4 \times 6}$. After the forward pass of SMPL with the regressed angles, joint positions $\mathbf{P} \in \mathbb{R}^{4 \times 3}$ are obtained. 

\paragraph{\textbf{Training}}

The proposed architecture is implemented in PyTorch and trained on an NVIDIA Tesla V100S GPU. The Adam optimizer is used with an initial learning rate of $1\times10^{-3}$ and a batch size of 512. The network is trained for 15 epochs. As described in Equation \ref{eq:loss}, the loss function $\mathcal{L}$  combines the mean absolute error (MAE) between predicted joint rotations and ground-truth rotations with the MAE computed on joint positions obtained via SMPL's forward pass, ensuring both rotational and positional accuracy.

\begin{equation}
\mathcal{L} = \lVert \mathbf{\Theta}_{pred} - \mathbf{\Theta}_{gt} \rVert_1
+  \lVert \mathbf{P}_{pred} - \mathbf{P}_{gt} \rVert_1,
\label{eq:loss}
\end{equation}

For training and validation, 10\% of the training data is randomly selected as a validation set. The test set is strictly separated from the training data and is used to evaluate generalization on unseen scenarios. Depending on the evaluation protocol described in Section~\ref{sec:eval_metrics}, the test set consists of either data from an unseen subject or from movements not present in the training corpus, ensuring that the model is assessed on truly novel conditions.

\section{Evaluation} \label{sec:5}

\subsection{Experiment Settings}

To train a robust motion-capturing model, it is crucial to acquire a dataset that captures a wide range of human lower-limb movements under realistic conditions. The experiment routine was therefore designed to systematically cover the principal physiological axes of motion, while also including natural, unconstrained activities to reflect daily human behavior. This ensures that the resulting dataset not only provides diverse input for model generalization but also represents the variability encountered in real-world wearable applications. The designed protocol includes isolated movements, workout exercises, and daily activities.

\begin{itemize}
    \item \textbf{Isolated Movements.}
    This phase includes four fundamental movements where a single degree of freedom is primarily active. Examples include knee flexion and hip abduction, which isolate the contribution of individual joints.
    
    \item \textbf{Workout Exercises.}
    The second phase introduces eight compound exercises involving multiple joints and axes of freedom. These movements, commonly encountered in standard gym routines, aim to capture richer kinematic variation and joint coordination patterns.
    
    \item \textbf{Daily Movements.}
    The final phase comprises four movements typical of daily life, such as stepping up, sitting down, and walking. In addition, each participant performs one improvised movement of their choice to capture subject-specific motion styles.
\end{itemize}

All phases except walking are executed in a single continuous session lasting approximately ten minutes. Participants complete two sessions of the workout phase, followed by a five-minute walking phase. Experiments are conducted indoors in a spacious room to allow full mobility. A smartphone with a monocular RGB camera is positioned to record each session for subsequent ground-truth extraction using the WHAM framework. Supplementary equipment, such as a step platform and chair, is provided where necessary. The complete list of the movements can be found under Table \ref{tab:movement_list}.

\begin{table}[]
\centering
\resizebox{0.9\columnwidth}{!}{%
\begin{tabular}{|l|l|l|}
\hline
\rowcolor[HTML]{C0C0C0} 
\textbf{Category}                    & \textbf{ID} & \textbf{Movement Name}           \\ \hline
                                     & 1           & Hip Extension/Flexion            \\ \cline{2-3} 
                                     & 2           & Hip Abduction/Adduction          \\ \cline{2-3} 
                                     & 3           & Hip Rotation                     \\ \cline{2-3} 
\multirow{-4}{*}{Isolated Movements} & 4           & Knee Extension/Flexion           \\ \hline
                                     & 5           & Squats                           \\ \cline{2-3} 
                                     & 6           & Lateral Lunges                   \\ \cline{2-3} 
                                     & 7           & Lateral Lunges with Knee Drive   \\ \cline{2-3} 
                                     & 8           & Curtsy Lunge                     \\ \cline{2-3} 
                                     & 9           & Rainbow Leg Lift with Knee Drive \\ \cline{2-3} 
                                     & 10          & Butterfly                        \\ \cline{2-3} 
                                     & 11          & Alternating Legs on Floor        \\ \cline{2-3} 
\multirow{-8}{*}{Workout Exercises}  & 12          & Bulgarian Split Squat            \\ \hline
                                     & 13          & Step Up/Down                     \\ \cline{2-3} 
                                     & 14          & Sit Up/Down                      \\ \cline{2-3} 
                                     & 15          & Walking                          \\ \cline{2-3} 
\multirow{-4}{*}{Daily Movements}    & 16          & Random Movements                 \\ \hline
\end{tabular}%
}
\caption{List of movements used in data collection experiments.}
\label{tab:movement_list}
\end{table}

\paragraph{\textbf{Procedure.}}
Participants are not instructed to strictly follow or mimic a predefined movement pattern. This intentional relaxation promotes natural motion and intra-participant variability. No fixed number of repetitions is prescribed, allowing variations in temporal duration. For asymmetric movements (e.g., hip flexion, hip abduction), participants are encouraged to balance repetitions between the left and right legs. The final movement in each session is fully improvised to capture individualized motion dynamics.

Each movement interval consists of 30 seconds of activity followed by 10 seconds of rest. During rest, participants tap three times on the right front-hip channel to mark the transition between movements, enabling the segmentation of each movement later. Both auditory and visual cues are provided to synchronize stages precisely.

\subsection{Data Collection}\label{sec:datacollection}
A total of 11 volunteers (6 males and 5 females) participated in the study, resulting in approximately 4 hours of recorded data. Each session includes recordings from the textile capacitive sensors and the video-based WHAM reference system. To mitigate slight mismatches in the sensor data resulting from the stochastic settlement time of the FDC2214 channels, the ground-truth data is interpolated to match the timestamps from DAU. All participants provided informed consent to take part in this preliminary pilot study. Ethical review and approval were waived for this study, as the collected data were fully anonymized and no identifiable visual information was shared or presented. Collected data was limited to sensor readings located on the textile layer of the pants. Standardized 3D body models fitted to video data were used as body pose reference to keep the data anonymized, and the 3D model was normalized by participants' tibia length. The data were exclusively used for developing and training deep learning models, without deriving medical interpretations or providing individual feedback.

During preliminary data analysis, the two "RearHip" channels were excluded due to intermittent disconnectivity issues, persistent throughout the majority of recordings. Additionally, 4 recordings were also discarded because of jittery or blurred camera footage, yielding a final set of 29 valid sessions.

After normalization of capacitance values to the range $[0,1]$, windows containing samples outside this interval (caused by intermittent sensor disconnections) were removed. Consequently, the total dataset size was reduced from 440{,}249 samples (approximately 4.1 hours) to 398{,}400 clean samples (approximately 3.7 hours). These processed data form the basis for model training and evaluation described in Section~\ref{sec:4}.

\subsection{Evaluation Metrics} \label{sec:eval_metrics}
To assess both reconstruction accuracy and deployability, we employ two cross-validation strategies and a set of spatial, angular, temporal, and computational metrics, consistent with prior pose estimation works~\cite{mocapose,seampose,avatarposer,Transpose,emdb}.

\paragraph{\textbf{Leave-One-Participant-Out (LOPO)}}
Given a single-size pair of pants, the relative alignment between the textile patches and anatomical landmarks may vary across individuals. Differences in height, limb length, and joint size alter the deformation patterns of the patches, resulting in different signal characteristics. For example, the knee patches of a shorter participant may align slightly above the patella, while for a taller one they may sit below, producing distinct capacitance responses. To evaluate how robust the model remains to unseen users, $11$-fold cross-validation approach is employed: $11$ models are trained, each of which excludes one participant for testing. This approach assesses how well the model adapts to unseen subjects with differing body shapes and garment fits.

\paragraph{\textbf{Leave-One-Exercise-Out (LOEO)}}
To evaluate the model's capacity to generalize to unseen motions, a $16$-fold cross-validation approach is employed again, this time excluding a unique movement from Table \ref{tab:movement_list}. The fixed duration of each experiment enables precise temporal segmentation and exclusion of each exercise. This assesses whether the learned temporal and spatial representations can extend to new motion types not observed during training.

Two primary metrics are used for quantitative evaluation.
\begin{itemize}
\item{\textbf{Mean Per Joint Angle Error (MPJAE):}}
MPJAE measures the average angular deviation between predicted and ground-truth joint rotations \cite{ludwig2025}. The model outputs 6D rotation representations, which are first converted into $3{\times}3$ rotation matrices. The geodesic distance between the predicted rotation $\hat{R}$ and ground truth $R$ is then calculated for MPJAE (Equation \ref{eq:mpjae}).

\begin{equation}
\text{MPJAE} = \frac{1}{4}\sum_{j=1}^{4} \arccos\!\left(\frac{\text{trace}\left(R_j^\top \hat{R}_j\right) - 1}{2}\right)
\label{eq:mpjae}
\end{equation}

where $J = 4$, is the number of joints, representing \texttt{left\_hip}, \texttt{right\_hip}, \texttt{left\_knee}, \texttt{right\_knee}.

\item{\textbf{Mean Per Joint Position Error (MPJPE):}}
MPJPE quantifies the Euclidean distance between predicted and ground-truth joint positions in 3D space for \texttt{left\_knee}, \texttt{right\_knee}, \texttt{left\_ankle}, and \texttt{right\_ankle}, as computed in Equation \ref{eq:mpjpe}.

\begin{equation}
\text{MPJPE} = \frac{1}{4}\sum_{j=1}^{4} \left\| \mathbf{p}_j - \hat{\mathbf{p}}_j \right\|_2
\label{eq:mpjpe}
\end{equation}

where $\mathbf{p}_j$ and $\hat{\mathbf{p}}_j$ denote the ground-truth and predicted 3D joint coordinates, respectively. These positions are obtained from the SMPL model’s forward kinematics given the predicted joint rotations.

\item \textbf{Jitter:}
To capture the temporal consistency and smoothness through the consecutive predictions of joint positions, we adopt the "Jitter" metric introduced in~\cite{Transpose} and widely employed in subsequent motion capture studies~\cite{emdb}. The metric corresponds to the time derivative of acceleration, or, namely, jerk~\cite{Flash1688}. In this work, jitter is computed from three successive discrete differences of the estimated joint positions obtained via the SMPL forward pass, as described in Equation \ref{eq:jerk}.

\begin{equation} \label{eq:jerk}
    \zeta[t] =  \left \lVert {\textbf{p}[t] - 3\textbf{p}[t-1] +3 \textbf{p}[t-2] - \textbf{p}[t-3] \over (\Delta t)^3}\right \rVert_2
\end{equation}

where $\mathbf{p}[t]$ denotes the 3D joint position at frame $t$ and $\Delta t$ is the sampling interval. Lower $\zeta$ values indicate smoother, more temporally consistent motion trajectories.

\end{itemize}

\paragraph{\textbf{Normalization and Scaling.}}
Because SMPL’s output is parameterized in a normalized space dependent on the body shape coefficients $\beta$, direct positional errors may not correspond to true metric distances across individuals. To ensure comparability, a subject-specific scaling is applied: the tibia length (distance between the knee and ankle joints) in the SMPL mesh is rescaled to match each participant’s measured leg length, collected during the experiments, ensuring the actual spatial error calculation for MPJPE and jitter.

\paragraph{\textbf{Embedded Metrics.}}
To evaluate the model’s suitability for embedded and portable deployment, the following computational metrics are additionally considered:
\begin{itemize}
    \item \textbf{Number of Parameters:} The total count of parameters, including both trainable and non-trainable weights, represents the model’s overall complexity.
    \item \textbf{Model Size:} The corresponding memory footprint of the parameters when stored in float32 representation indicates the storage requirements for deployment.
    \item \textbf{Floating Point Operations (FLOPs):} The total number of arithmetic operations (multiplications and additions) required for a single forward pass of the network, serving as an estimate of computational cost.
\end{itemize}
These embedded metrics collectively provide a practical measure of the architecture’s deployability on resource-constrained edge devices, complementing the reconstruction accuracy metrics described above.

\subsection{Embedded Implementation} \label{sec:emb_impl}
To evaluate the real-time performance of the proposed deep learning model on an edge platform, an Android-based smartwatch application was developed in Kotlin. The trained models were converted to TensorFlow Lite (TFLite) format and embedded directly into the application. The target device is a commercial TicWatch Pro~3~GPS, developed by Mobvoi \footnote{https://www.mobvoi.com/us/pages/ticwatchpro3gps}, equipped with four ARM Cortex-A53 cores, one ARM Cortex-M0 core, and an Adreno~504 GPU.

No quantization or pruning techniques were applied to the models; all computations were executed in float32 precision to enable a fair runtime comparison between different architectures without altering their representational fidelity. The trained models were deployed on the smartwatch to compare their real-time on-device inference latency.

\begin{figure}[h]
  \centering
  \includegraphics[width=\columnwidth]{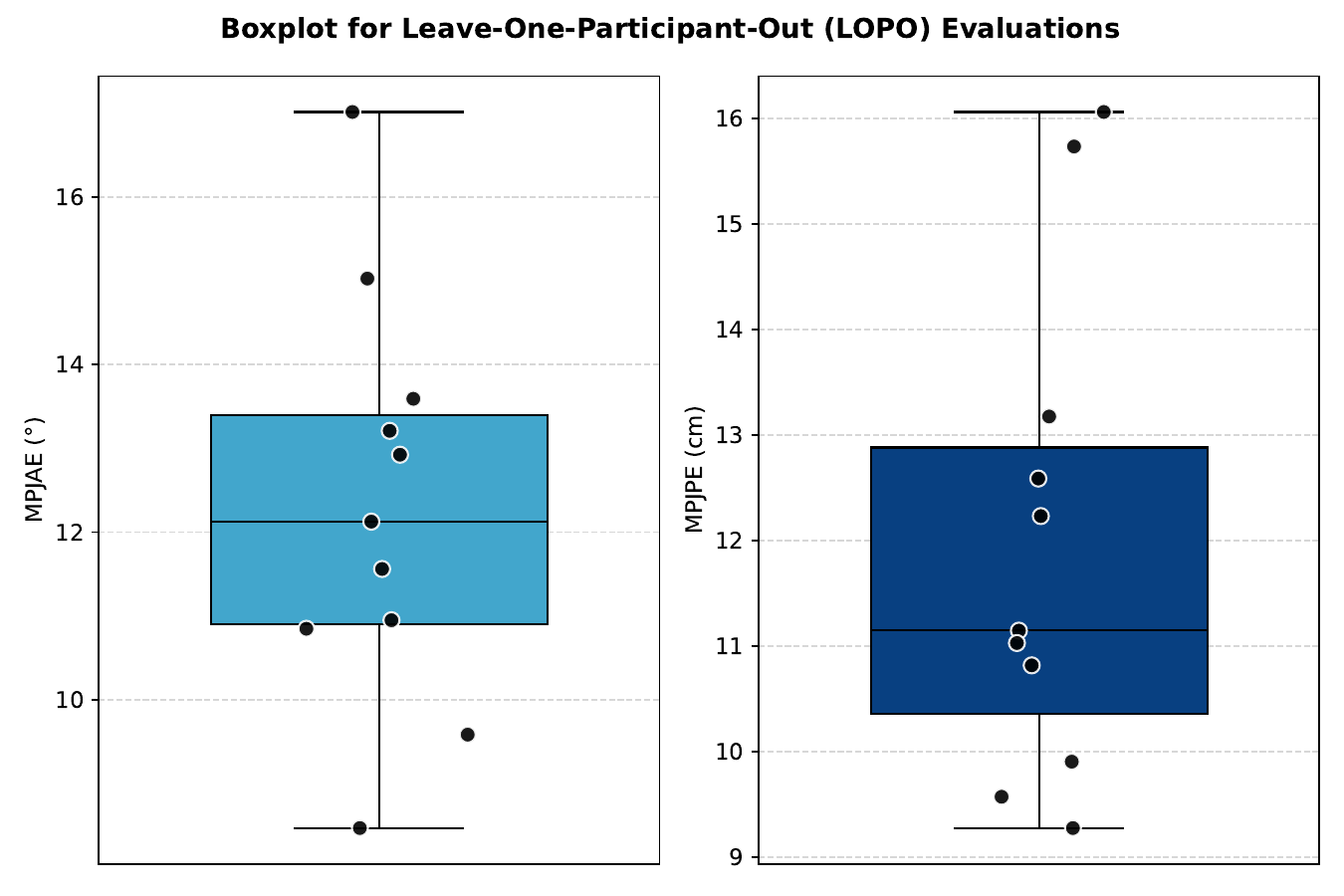}
  \caption{Results of Leave-One-Participant-Out evaluation with our proposed CNN2D+Transformer architecture.}
  \label{fig:lopo}
\end{figure}

\subsection{Performance}

\paragraph{\textbf{Unseen User Performance}}
Eleven models trained to assess unseen user performance, using LOPO protocol and the resulting metrics, averaged over all lower-body joints for each participant, are presented in Figure~\ref{fig:lopo}. Across all folds, the proposed model achieves a mean position error of $11.96~(\pm 2.19)$~cm and a mean rotation error of $12.30~(\pm 2.32)^{\circ}$.

As no extensive subject-specific calibration is performed (except for the baseline channel subtraction), these results reflect the system’s inherent ability to generalize across individuals with varying body shapes and garment fits. Despite the variability in fit and signal alignment between participants, the variance across subjects remains small, indicating stable model behavior under unseen conditions.

\begin{figure}[h]
  \centering
  \includegraphics[width=\linewidth]{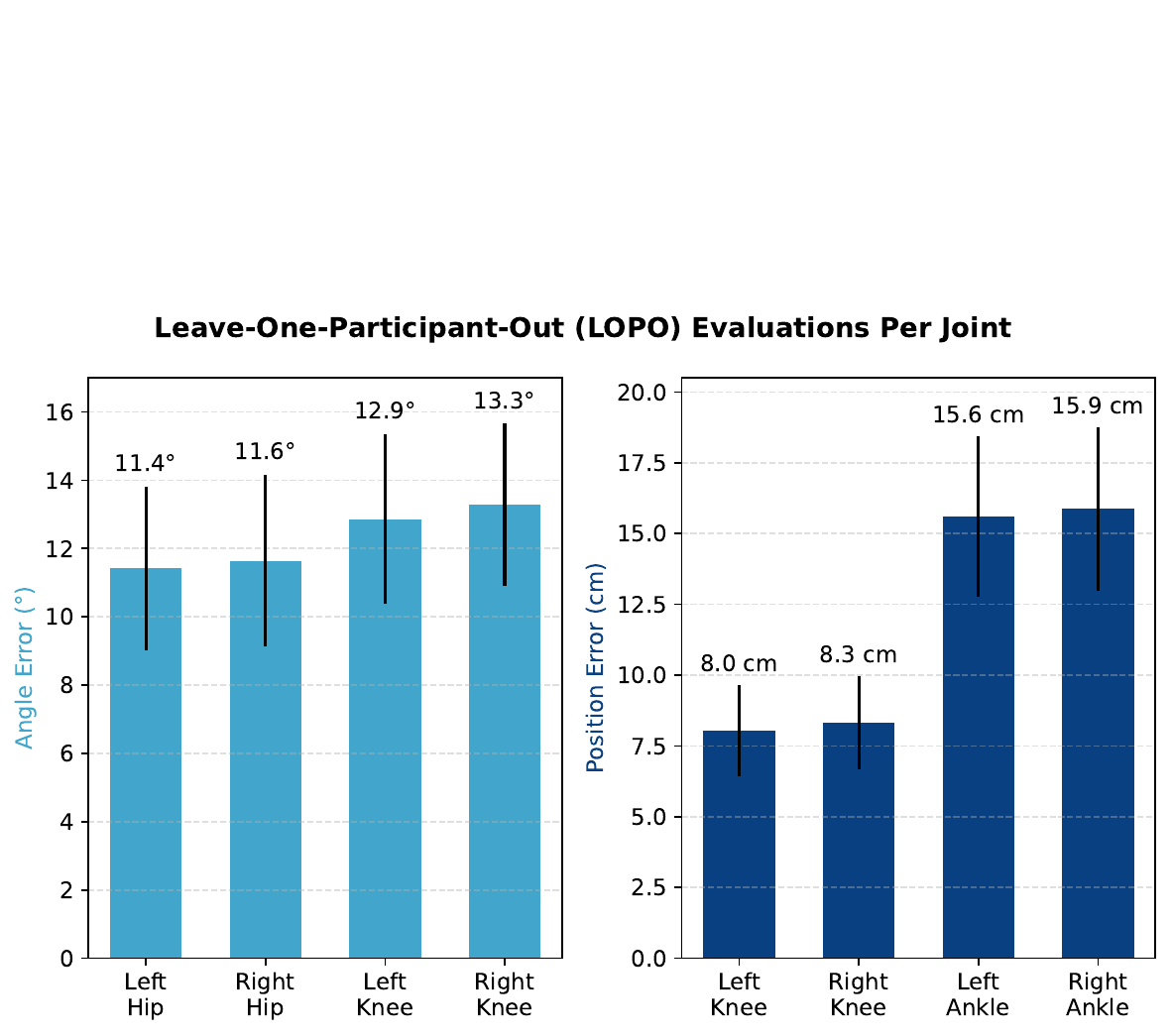}
  \caption{MPJPE and MPJAE Metrics Averaged Per Joint. Error bars represent the interparticipant standard deviation for joint errors.}
  \label{fig:per_joint}
\end{figure}

From a joint-level perspective (Figure~\ref{fig:per_joint}), the average rotation error for the hip joints is $11.5^{\circ}$, while the knee joints exhibit $13.1^{\circ}$ of rotation error, with a slight asymmetry between left and right limbs. The corresponding mean position error for the knees is $8.2$~cm. As expected from forward kinematics, rotation errors accumulate as the body moves away from its center, resulting in the highest positional error at the ankle joints, with an average error of 15.8 cm.  

This mild asymmetry between limbs can be attributed to differences in signal integrity across left and right channels, as well as inconsistencies in the data collection process, as participants were not instructed to perform a fixed number of repetitions, which introduced uneven coverage between sides.

\begin{figure*}[h]
  \centering
  \includegraphics[width=\textwidth]{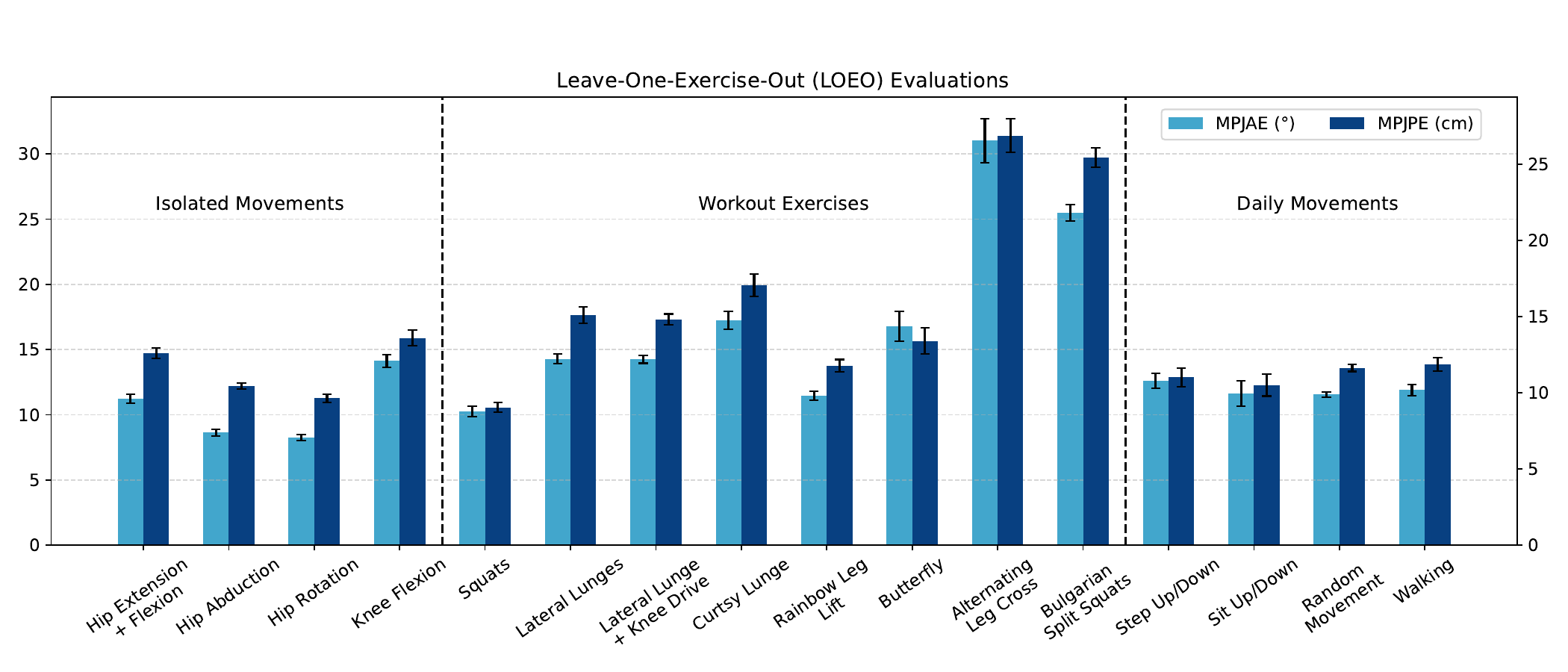}
  \caption{Results of Leave-One-Exercise-Out evaluation with CNN2D+Transformer architecture. Grouped by the type of activity groups as stated in Table \ref{tab:movement_list}: Isolated Movements, Workout Exercises, Daily Movements. Error lines represent the standard deviation of experimental iterations.}
  \label{fig:loeo}
\end{figure*}

\begin{table*}[]
\centering
\resizebox{\textwidth}{!}{%
\begin{tabular}{cccccccc}
\rowcolor[HTML]{C0C0C0} 
\textbf{Architecture}             & \textbf{\begin{tabular}[c]{@{}c@{}}MPJPE\\ (cm)\end{tabular}} & \textbf{\begin{tabular}[c]{@{}c@{}}MPJAE\\ ($\circ$)\end{tabular}} & \textbf{\begin{tabular}[c]{@{}c@{}}Jitter\\ ($10m \times s^{-3}$)\end{tabular}} & \textbf{\begin{tabular}[c]{@{}c@{}}Number of\\ Parameters (K)\end{tabular}} & \textbf{\begin{tabular}[c]{@{}c@{}}Model Size\\ (MB)\end{tabular}} & \textbf{\begin{tabular}[c]{@{}c@{}}FLOPs\\ (M)\end{tabular}} & \textbf{\begin{tabular}[c]{@{}c@{}}Inference Time\\ (ms)*\end{tabular}} \\ \hline
BiLSTM-based \cite{seampose}      & 12.13 ($\pm$0.23)                                             & 12.46 ($\pm$0.21)                                                & 23.32 ($\pm$0.55)         & 2,490                                                                       & 9.5                                                                & 567.96                                                       & 860.20 ($\pm$7.57)                                                     \\
CNN2D+CNN1D-based \cite{mocapose} & 12.54 ($\pm$0.34)                                             & 13.00 ($\pm$0.32)                                                & 32.00 ($\pm$0.48)         & 334                                                                         & 1.27                                                               & 121.46                                                       & 60.27 ($\pm$3.14)                                                      \\
Conv2D + Transformer (ours)       & \textbf{11.96 ($\pm$0.22)}                                    & \textbf{12.30 ($\pm$0.22)}                                       & \textbf{21.6 ($\pm$0.40)} & \textbf{106}                                                                & \textbf{0.42}                                                      & \textbf{30.54}                                               & \textbf{23.68 ($\pm$1.46)}                                             \\ \hline
\multicolumn{8}{l}{\small * Executed and averaged for 100 iterations.} \\
\end{tabular}%
}
\caption{The Metrics for Comparison of Three Architectures.}
\label{tab:comparative_results}
\end{table*}

\paragraph{\textbf{Unseen Exercise Performance}}
It is impractical to capture every possible lower-body motion during data collection; therefore, the model’s ability to generalize to new movements was evaluated using the LOEO strategy. The aggregated results of 16 models are shown in Figure~\ref{fig:loeo}. The model achieves an average position error of $14.06~(\pm 5.02)$~cm and an average rotation error of $14.42~(\pm 5.83)^{\circ}$ under this protocol.  

When grouped by movement categories introduced in Table~\ref{tab:movement_list}, isolated movements yield the lowest rotation error of $10.56^{\circ}$, followed by daily movements with $11.92^{\circ}$, and workout exercises with $17.60^{\circ}$. These results align with expectations: isolated movements involve single degrees of freedom and simple joint trajectories, whereas workout exercises engage multiple joints simultaneously, increasing kinematic complexity.

At a finer level of granularity of movements, the lowest rotation error is observed during the "Hip Rotation" movement ($8.25^{\circ}$), while the highest occurs in "Alternating Leg Cross" ($31.01^{\circ}$). The latter involves floor contact and complex inter-limb coordination, contributing to higher intra-participant variability. Participants often executed this exercise inconsistently, as the visual reference for the movement was much more challenging to mimic precisely compared to other movements.

Overall, errors under the LOEO protocol are slightly higher than those observed in the LOPO setting. In the unseen-user case, the model learns to accommodate inter-subject variations within a fixed movement corpus. In contrast, unseen-movement evaluation tests the model’s ability to extrapolate to novel joint coordination patterns not encountered during training. These findings suggest that extending the movement corpus with additional motion types (such as dancing and yoga) could further enhance the generalization capability of the learned representation.

\paragraph{\textbf{Comparison with Literature.}}
A direct quantitative comparison with existing studies on loose-fitting textile capacitive sensing, such as SeamPose~\cite{seampose} and MocaPose~\cite{mocapose}, is not feasible due to the absence of publicly available datasets and the use of custom garments and circuitry in each study. Moreover, both works focus on the upper body, whereas this study is the first to target the lower body using loose-fitting textile sensing. Compared to the upper body, the lower body involves fewer but more dynamic joints (hips and knees, whose motion spans a broader range of rotation and translation). In contrast, upper-body joints such as the neck and shoulders remain relatively stationary during most movements.

MocaPose reports a mean per-joint position error (MPJPE) of $8.8$~cm, and SeamPose reports $8.6$~cm. Our model achieves an MPJPE of $11.96$~cm over four lower-body joints, for unseen users. Considering that the hip and knee joints are functionally analogous to the elbow and wrist joints in upper-body studies, which also report the highest errors in both~\cite{mocapose,seampose}, with 9.6~cm and 9.1~cm for elbows and 13.6~cm and 16.5~cm for wrists, respectively, the results demonstrate that the proposed framework achieves comparable reconstruction accuracy under a more challenging sensing configuration.

In addition to performance comparison, we evaluated the architectures proposed in previous works under our experimental setup. The convolutional hybrid model (CNN1D+CNN2D) from MocaPose and the bidirectional LSTM-based model from SeamPose were both reimplemented with input and output dimensions adapted to our system. The input window length was set to 120 frames, longer than the 32 and 96 samples used in~\cite{mocapose} and~\cite{seampose}, respectively, to better capture temporal dependencies in motion. Using the same training configuration and the LOPO evaluation, results summarized in Table~\ref{tab:comparative_results} show that our proposed Transformer-based architecture slightly outperforms both baselines across all metrics. This improvement can be attributed to the Transformer's ability to model long-range temporal dependencies, which are crucial when interpreting smooth, continuous motion from capacitive time-series data.

\paragraph{\textbf{Embedded Implementation Comparison.}}
An additional objective of this study is to ensure that the proposed model remains computationally efficient for embedded deployment. For this, we compare the parameter count, floating-point operations (FLOPs), and real-time inference time of all three architectures following the methodology described in Section~\ref{sec:emb_impl}. The results, presented in Table~\ref{tab:comparative_results}, highlight the compactness of our approach.

The CNN2D+CNN1D model contains approximately $2.15\times$ more parameters than our proposed architecture, while the BiLSTM-based model is $22.49\times$ larger. In terms of computational complexity, our model requires roughly $3\times$ fewer FLOPs than the CNN-based model and $17.6\times$ fewer than the BiLSTM counterpart. These differences are significant for memory and energy-constrained edge devices.

Moving from theoretical to empirical evaluation, runtime measurements on the smartwatch demonstrate that our deep learning architecture achieves an average forward-pass inference latency of $23.68$~ms per frame, corresponding to approximately 42.2~frames per second (FPS). In comparison, the CNN2D+CNN1D model requires $60.27$~ms ($\sim$16.6~FPS), while the BiLSTM-based model exhibits an impractically high latency of $860.20$~ms ($\sim$1.2~FPS). We target a processing rate consistent with our monocular camera’s frame rate ($\sim$25 - 30 FPS) to ensure visually smooth reconstruction. Since these measurements reflect only the model’s forward-pass inference, our approach is the only one to meet this frame-rate target, leaving sufficient room for additional pre- and post-processing within the application pipeline.

Considering both reconstruction performance and computational efficiency, our Transformer-based model achieves a favorable trade-off. Although its performance improvements over prior works are moderate, $4.63\%$ and $5.38\%$ better MPJPE and MPJAE metrics than MocaPose, and $1.24\%$ and $1.28\%$ improvements over SeamPose, its dramatically reduced parameter count and inference latency make it substantially more suitable for deployment in embedded, real-time wearable systems. 

Furthermore, extending beyond smartwatches to more resource-constrained platforms such as microcontrollers for any possible future deployments, our model remains within practical limits: its 416.6~KB memory footprint fits comfortably within the 1~MB flash and 512~KB RAM of the VersaSens platform~\cite{versasens}, whereas the other two models exceed these constraints. These results validate the proposed architecture as a compact yet high-performing solution for textile-based motion capture.

\begin{figure}[h]
  \centering
  \includegraphics[width=\linewidth]{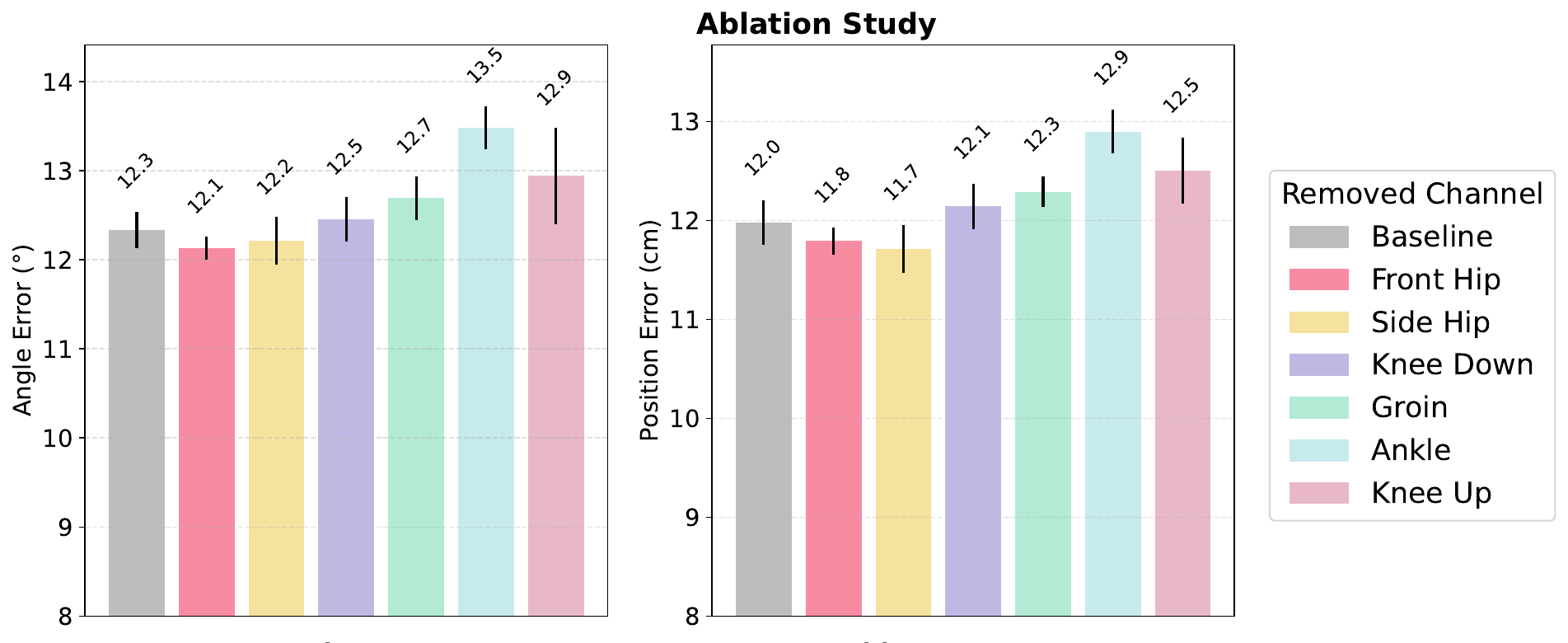}
  \caption{Ablation Study for Channel Importance. The effect of removing channels one-by-one is presented for both MPJAE (left) and MPJPE (right) metrics using LOPO approach, with standard deviations provided.}
  \label{fig:ablation}
\end{figure}

\begin{figure*}[h]
  \centering
  \includegraphics[width=\textwidth]{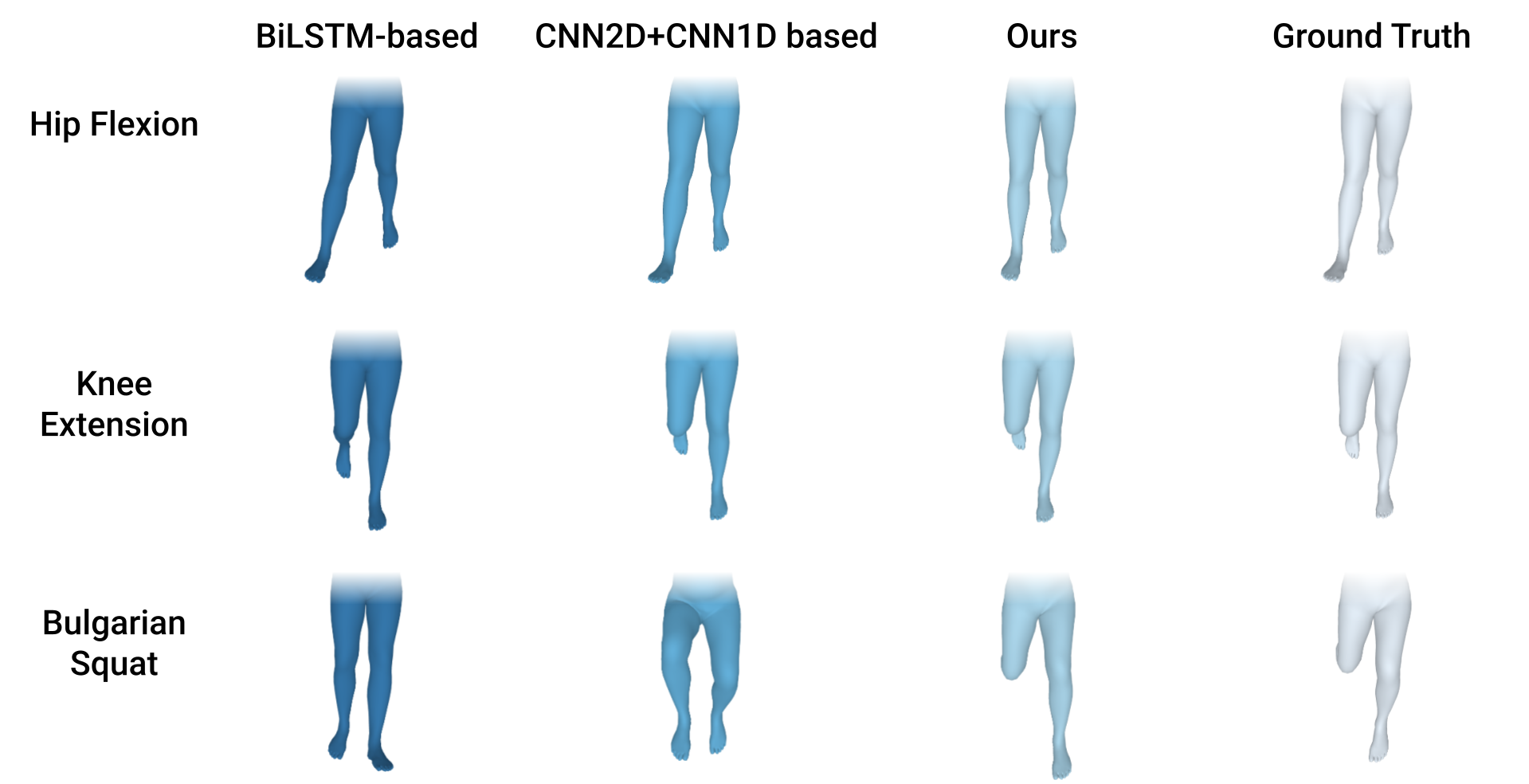}
  \caption{SMPL Visualization of Three Movements. The ground truth is compared with three architectures evaluated. Hip Flexion: All of the models perform similar reconstruction. Knee Extension: We observe slight mismatch with the timing of the reconstructed motion on BiLSTM-based model. Bulgarian Squat: Some windows create random motion for BiLSTM-based and CNN2D+CNN1D-based models.}
  \label{fig:smpl_Viz}
\end{figure*}

\paragraph{\textbf{Channel Ablation Study}}
To evaluate the contribution of each sensing channel, a channel ablation study was conducted. As previously discussed, the proposed system is over-constrained, containing more sensing channels than the actual degrees of freedom of the lower body. Nevertheless, identifying redundant channels is valuable for simplifying the system design, improving cable management, and reducing potential crosstalk, since deactivating channels lowers the number of active FDC2214 chips/channels in the final implementation.

Given the symmetric configuration of the garment, channels were removed in symmetric pairs, resulting in six ablation combinations. As illustrated in Figure~\ref{fig:ablation}, the removal of individual channel pairs produced no significant degradation in overall model performance. The largest increase in error was observed when the "Ankle" channel was excluded, with approximately 1~cm increase in position error and 1° increase in angular error.

Although the ankle region exhibits minimal local deformation along the tibia, this channel’s relevance likely arises from its displacement response rather than direct strain sensing. When the knee flexes, inter-participant differences in pant fit and patch alignment cause the knee channels to register variable signal distributions. To mitigate this, two vertically offset knee patches, "KneeUp" and "KneeDown", were included (as detailed in Section~\ref{sec:wearable_design}) to help the model infer alignment implicitly. However, unlike these channels, the "Ankle" patch remains consistently positioned along the tibia, its displacement governed primarily by knee flexion and extension. This stability likely explains its disproportionate importance to accurate angle estimation.

The removal of "Front Hip" and "Side Hip" yields slightly lower error than the baseline. Nonetheless, it is worth noting that the current prototype exhibits non-ideal signal integrity. During experiments, minor crosstalk and wire management artifacts introduced noise across several channels, with varying magnitudes between the left and right counterparts, as well as occasional connectivity problems. Consequently, while the ablation results suggest promising robustness, they should be interpreted cautiously. A refined prototype with improved shielding and cable routing is required to draw definitive conclusions about channel redundancy.

\begin{table*}[]
\centering
\begin{tabular}{ccccccccc}
\rowcolor[HTML]{C0C0C0} 
\textbf{}              & \textbf{Metric} & \textbf{Mean}  & \textbf{Left Hip} & \textbf{Right Hip} & \textbf{Left Knee} & \textbf{Right Knee} & \textbf{Left Ankle} & \textbf{Right Ankle} \\ \hline
                       & MPJPE (cm)      & \textbf{11.96} &             -      &           -         & 8.03               & 8.33                & 15.60               & 15.86                \\
\multirow{-2}{*}{LOPO} & MPJAE ($\circ$)   & \textbf{12.30} & 11.42             & 11.64              & 12.86              & 13.27               &            -         &             -         \\ \hline
                       & MPJPE (cm)      & \textbf{14.06} &        -           &         -           & 9.17               & 9.62                & 18.69               & 18.72                \\
\multirow{-2}{*}{LOEO} & MPJAE ($\circ$)   & \textbf{14.42} & 13.02             & 13.42              & 15.43              & 15.80               &          -           &              -        \\ \hline
\end{tabular}
\caption{The final metrics of our study, evaluated for LOPO and LOEO approaches, with corresponding joint-based errors.}
\label{tab:lopo_loeo_perjoint}
\end{table*}

\section{Limitations and Discussion} \label{sec:6}

\subsection{Prototype Quality}
The primary limitation affecting system performance originates from the prototype’s signal integrity and connectivity. Post-experiment analysis revealed that several channels suffered from unstable connections and noisy signals. Although embedding copper wires within the seams offered a neat cabling solution, the use of DuPont connectors between the garment and the Data Acquisition Unit (DAU) introduced intermittent disconnections. Given the dynamic movements of participants, small displacements or stretching of the pants often disrupted contact, resulting in missing data and the eventual removal of the "RearHip" channels. Future iterations should adopt more robust connectors, such as magnetic clips proposed in~\cite{seamfit}, to ensure stable electrical contact even under motion.

With fourteen total channels (seven per leg), all four FDC2214 chips were simultaneously active. To minimize cross-talk between chips, sequential sampling was employed, activating one chip at a time rather than parallel sampling. This design choice preserved signal integrity but reduced the effective integration time per channel, limiting the achievable sampling resolution. Insights from the channel ablation study suggest that future versions could reduce the number of channels to improve both per-channel resolution and overall signal stability, while also simplifying cabling and lowering cross-talk within chips.

\subsection{Ground Truth Limitations}
Beyond the input quality, reliable ground-truth joint data are critical. Because marker-based motion capture is infeasible for loose-fitting garments, where markers move relative to the body due to the loose garment, we relied on monocular pose estimation via the WHAM framework. However, WHAM itself is trained using another dataset and inherits an average positional error of approximately 6.25~cm (averaged MPJPE across three datasets used in the study) \cite{wham}. This inherent uncertainty propagates through our system, introducing a slight deviation between the predicted and actual joint angles.

Further limitations arose from the recording setup. WHAM occasionally failed to distinguish the left and right legs during occlusions or when participants faced away from the camera. This was likely due to the single-camera configuration and the black pants worn by participants, which reduced the contrast of leg contours. Future experiments should employ lighter-colored garments and multi-view camera setups to alleviate occlusion effects and improve 3D reconstruction consistency.

A final issue concerns synchronization between the CSB data and video frames. The DAU sampled at approximately 30.15~Hz, while the smartphone camera operated at 29.5–29.99~FPS. Although we interpolated the SMPL parameters to align with sensor timestamps (see Section \ref{sec:datacollection}), even minor temporal offsets caused noticeable misalignment between the signal and corresponding ground-truth labels. Since the deep learning model learns mappings between sensor inputs and temporally aligned joint angles, such desynchronization directly impacts performance. A more precise synchronization method should be implemented in future experiments.

\subsection{Experimental Limitations}
The dataset was collected from 11 participants, providing a reasonable but limited range of body morphologies. Expanding the participant pool would enhance model generalization and robustness, particularly for users who are unseen. Additionally, the difference in performance between the "LOPO" and "LOEO" evaluations suggests that the model generalizes better to new users than to unseen movements. To enhance coverage of motion variability, the movement corpus (currently consisting of 16 daily and exercise-based movements) should be extended to include more diverse activities, such as dance or sports-specific actions, which would enrich the model’s learning space.

\subsection{Scalability and Producibility}
For large-scale or repeatable production, it is essential to ensure that data and models remain transferable across garments. Minor discrepancies in patch dimensions, material properties, or placement can significantly affect capacitance readings, potentially invalidating previously trained models. In future work, custom-designed textile templates should be used to standardize patch positioning and sewing patterns, enabling consistent manufacturing of multiple prototypes and ensuring cross-compatibility between collected datasets.

Another practical consideration is garment washability. As the system targets motion monitoring and sports applications, the ability to withstand washing is crucial. While the current design allows detachment of the DAU for safety, the conductive textiles and wiring still need to be tested under water exposure and chemical stress. Exploring alternative conductive materials or integrating stitched patch connections could improve long-term durability and enable routine cleaning without degradation.

\subsection{Future Work}
VersaPants represents the first attempt to perform lower-body motion capture using loose-fitting textile capacitive sensing. While this study validates the feasibility of this approach, numerous directions exist for future enhancements. Incorporating anthropometric features could help personalize predictions by providing the model with implicit cues about individual body geometry, which strongly influences sensor response. Beyond local joint orientation estimation, future systems could also infer the global pose and translation of the body by fusing data from complementary sensors, such as the IMU integrated into the VersaSens board or the smartwatch used for embedded inference. Also, the multimodal nature of the VersaSens platform enables the possibility of sensor fusion with other modalities, such as EMG, bioimpendace, etc. Finally, moving toward a more sustainable version of VersaPants by replacing the garment with recyclable, eco-friendly textiles and transitioning from silicon-based electronics to sustainable e-textile circuitry based on Organic Electrochemical Transistors (OECTs) and other bio-derived components could provide a more environmentally responsible solution aligned with future sustainability goals.

\section{Conclusion} \label{sec:7}
This work introduces VersaPants, a novel system for lower-body motion capture using loose-fitting textile capacitive sensing and a compact deep learning model. By combining a wearable prototype, a cost-efficient data acquisition pipeline, and a lightweight Transformer-based architecture, the system enables joint angle regression in the SMPL body model space from simple capacitance signals. Experimental results with 11 participants and 16 diverse movements demonstrate that VersaPants achieves competitive accuracy with state-of-the-art wearable approaches while maintaining a significantly smaller computational footprint, supporting real-time embedded deployment.

While current results validate the feasibility of textile-based capacitive sensing for motion capture, several limitations remain, including signal integrity, synchronization precision, and dataset diversity. Future work will focus on improving the prototype’s hardware robustness, refining ground-truth quality through multi-view capture, and exploring multimodal sensor fusion. The findings of this study pave the way for smart garments for lower-body human motion analysis and interactive applications.

\bibliographystyle{ACM-Reference-Format}
\bibliography{main}
\end{document}